\documentclass[twocolumn, prl]{revtex4}
\usepackage{graphicx}
\usepackage{epsfig}
\begin{document}
\title{Homogeneous Fermion Superfluid with Unequal Spin Populations }
\author{Tin-Lun Ho and Hui Zhai}
\affiliation{Department of Physics, The Ohio State University, Columbus, OH}
\date{\today}
\begin{abstract}
For decades, the conventional view is that an s-wave BCS superfluid can not support uniform spin polarization due to a gap $\Delta$ in the quasiparticle excitation spectrum. We show  that this is an artifact of the dismissal of quasiparticle interactions $V_{qp}^{}$ in the conventional approach at the outset.  Such interactions can cause triplet fluctuations in the ground state and hence non-zero spin polarization at 
``magnetic field" $h<\Delta$.
The resulting ground state is a pairing state of quasiparticles on the ``BCS vacuum".  
For sufficiently large $V_{qp}$, the spin polarization of at unitarity has the simple form $m\propto \mu^{1/2}$. 
Our study is motivated by the recent experiments at Rice\cite{Rice} which found evidence of a homogenous superfluid state with uniform spin polarization. 
\end{abstract}
\maketitle

{ \em What happens to an s-wave BCS superfluid when the number of up and down spins become unequal?}
 For 50 years since the  invention of BCS theory of superconductivity, this  fundamental question remains unsettled.  Recent experiments on degenerate Fermi gases of $^{6}$Li at MIT\cite{MIT}and at Rice University\cite{Rice}, however, have shed lights on this long standing problem. 

According to mean field theory, the BCS ground state is highly resistant to chemical potential difference $\delta \mu \equiv 2 h$ between the two spin populations. The system can not develop spin asymmetry (i.e. with zero susceptibility) unless $h$ is of the order the energy gap $\Delta$.  It is found that for $h>\Delta/\sqrt{2}$, the BCS state is unstable against the spin polarized normal state\cite{FFLO}. Exactly how the BCS state evolves as $h$ increases, or more relevant to current experiments, as spin polarization increases, is a subject  of controversy.   
There have been many proposals. The most famous one is the so-called FFLO (Flude, Farrel, Larkin and Ovchinikov) state\cite{FFLO}; which is a pairing state with an oscillatory gap along a specific spatial direction. 
Recent studies of strongly interacting atomic Fermi gases\cite{Recent}, however, show that the 
FFLO state can exist only in an insignificant range of spin polarization, and that any amount of polarization will cause the gas  to phase separate into regions of unpolarized BCS state and spin polarized normal state. 
In the following,  to make contact between atomic Fermi gases and electrons in metals, 
we shall refer to the difference in spin densities $(n^{}_{\uparrow} - n^{}_{\downarrow})$ of a Fermi gas simply  as ``magentization" $m$, and  $h$ simply as ``magnetic field". 

In both Ref.\cite{MIT,Rice}, the density profiles of different spin populations of $^{6}$Li Fermi gas in the strongly interacting regime have been measured. The MIT data are for $P>0.1$ ($P=(n_{\uparrow} - n_{\downarrow})/(n_{\uparrow} + n_{\downarrow})= m/n$)  taken after the gas is released from the trap. The deduction of original spin density requires the understanding of the expansion dynamics of a strongly interacting Fermi gas -- an intriguing problem yet to be worked out.  The measurements of the Rice group were performed {\em in situ}, including polarization $0<P<0.1$ (with accuracy 0.03). Although there are no explicit determination of temperature, the experimental conditions are similar to those in earlier experiments in the superfluid phase. It is then reasonable to think that the system is a superfluid.  While both groups have observed similar spin density profiles at larger polarization, the Rice group also has found evidence of {\em uniform spin density when spin polarization $P< 0.1$}.  This is a highly surprising because such a state is forbidden 
within {\em mean field} BCS treatment.  (Although a recent Monte Carlo calculation suggests some features of such a state, the evidence is very indirect that it is hard to make a strong case\cite{Carlson}).   While more experiments are needed to further confirm this finding,
it raises the fundamental question of whether  there is any mechanism {\em at all} to allow Fermion superfluids to accommodate uniform magnetization. 
 

In this paper,  we show that quasi-particle interactions (due to interactions left over when reducing the actual Hamiltonian to the BCS form) is a natural mechanism to generate triplet spin fluctuations. 
These fluctuations will lead to a {\em non-zero response} to ``magnetic field" $h$, and hence a {\em homogenous superfluid with uniform spin polarization}. To realize this true ground state mathematically, 
however,  requires taking the kind of conceptual steps forward as in BCS theory.  It is  well known that due to number conservation, BCS state cannot be obtained by performing perturbation theory on the normal state. It can only be realized by constructing a coherent state in number space to allow pair fluctuations.  Likewise, due to the spin conservation, the true ground state in non-zero magnetic field can not be realized by performing perturbation of quasi-particle interactions on the BCS state. 
It can only be obtained by considering {\em coherent states in spin space allowing spin fluctuations}. 

{\bf A. The natural emergence of triplet excitations:} Let us first ask a simple question: What would be the ground state of the BCS Hamiltonian if we insist that the system has a very small spin, (say $S=1$), and with zero total momentum. The answer is to create a triplet pair with zero momentum on the  BCS ground state, $c^{\dagger}_{\bf k\uparrow}c^{\dagger}_{\bf -k\uparrow}|BCS\rangle$, where  
\begin{equation}
|BCS\rangle = \prod_{\bf k} {\cal P}_{\bf k}|0\rangle, \,\,\,\,\,\,\, 
{\cal P}_{\bf k}= u_{\bf k}^{} + v_{\bf k}^{} c^{\dagger}_{\bf -k\downarrow}  c^{\dagger}_{\bf k\uparrow}. 
\label{Pk} \end{equation}
${\cal P}_{\bf k}^{}$ creates a coherent state of pairs in the number space,  $u_{\bf k}^{}$ and $v_{\bf k}^{}$ are the coherence factors. Since $|BCS\rangle$  is annihilated by the quasi-particle operators $\alpha_{\bf k} = u_{\bf k}^{} c^{}_{\bf k\uparrow} + v_{\bf k}^{}c^{\dagger}_{-\bf k \downarrow}$, and $\beta_{\bf -k} = - v_{\bf k}^{} c^{\dagger}_{\bf k\uparrow} + u_{\bf k}^{} c^{}_{\bf -k\downarrow}$, we have 
\begin{equation}
c^{\dagger}_{\bf k\uparrow}c^{\dagger}_{\bf -k\uparrow}|0\rangle_{\bf k}^{}|0\rangle_{\bf -k}^{}
=  \alpha^{\dagger}_{\bf k} \alpha^{\dagger}_{\bf -k}( {\cal P}_{\bf k}^{}   {\cal P}_{\bf - k}^{}    |0\rangle). 
\label{triplet} \end{equation}
In other words, creating triplet excitations from the real vacuum is equivalent to creating quasi-particle pairs from the ``BCS vacuum".  Since attractive interactions cause fermion pairing, attractive interactions between quasiparticles will cause pairing between them. 

{\bf B1. Interaction between Quasi-particles: }  The Hamiltonian of atomic gases is $H = T + V$, where $T =  \sum_{\bf k} $ $[ (\xi_{\bf k} -h)c^{\dagger}_{\bf k\uparrow} c^{}_{\bf k\uparrow} + (\xi_{\bf k} + h)c^{\dagger}_{\bf k\downarrow} c^{}_{\bf k\downarrow}]$, $\xi_{\bf k}^{} = \epsilon_{\bf k}^{}-\mu$, $\epsilon_{\bf k} = \hbar^2 k^2/(2M)$, and $V$ is pseudo-potential for two-particle scattering\cite{pseudo}.  $V$ conserves both number and spin.  In the BCS approach, one considers a ground state $|BCS\rangle$ (eq.(\ref{Pk})) that allows fluctuation of singlet pairs. This leads to the reduced Hamiltonian $H_{BCS}^{} = 
T +  \sum_{\bf k}^{}[ \Delta c^{\dagger}_{\bf k\uparrow}c^{\dagger}_{\bf -k\downarrow} + h.c.]$, 
where $\Delta$ is determined self consistently as   
$\Delta = g  \delta(r) \partial_{r}^{}  [ r \langle \psi_{\downarrow}^{} ({\bf R} -{\bf r}/2)
\psi_{\uparrow}^{} ({\bf R} + {\bf r}/2)\rangle ] $ \cite{Castin}, $g=4\pi\hbar^{2} a_{s}^{}/M$, and $a_{s}^{}$ is the s-wave scattering length. Explicitly, we have 
\begin{equation}
1 = g\int \frac{ {\rm d}^{3}k}{(2\pi)^3}\left( \frac{1}{2E_{\bf k}} -  \frac{1}{2 \epsilon_{\bf k}}\right), 
\label{BCSgap} \end{equation}
where $E_{\bf k} = \sqrt{(\epsilon_{\bf k}-\mu)^2 + \Delta^2}$.
We then have 
\begin{equation} 
H_{BCS} = \sum_{\bf k} \left[  (E_{\bf k}-h)\alpha^{\dagger}_{\bf k} \alpha_{\bf k}^{}
+  (E_{\bf k} + h)\beta^{\dagger}_{\bf k} \beta^{}_{\bf k} \right] + E_{BCS}^{},
\end{equation}
where $E_{BCS}^{}=\sum_{\bf k} (\xi_{\bf k} - E_{\bf k})$ is the energy of the BCS state
(eq.(\ref{Pk}))\cite{co-factor}.
The fact that  eq.(\ref{BCSgap}) and $E_{BCS}^{}$ are independent of $h$ means the 
$|BCS\rangle$ state is rigid against spin polarization, since $m = - \partial (E_{BCS}^{}/\Omega)/\partial h = 0$ for $h<\Delta$, where $\Omega$ is the volume of the system.

To illustrate the essential physics of quasi-particle interaction,  we consider the following model
interaction $V_{qp}$, (its origin will be discussed in section {\bf C}): 
\begin{equation}
V_{qp} = - \frac{1}{\Omega}\sum_{\bf p, p'}^{} W({\bf p, p'}) \Gamma^{\dagger}_{\bf p} \Gamma^{}_{\bf p'}, 
\,\,\,\,\,\,
\Gamma_{\bf p'} = \alpha^{\dagger}_{\bf p'} \alpha_{\bf -p'}^{\dagger}  - \beta^{}_{\bf -p'} \beta^{}_{\bf p'}, 
\label{Vqp} \end{equation}
where $W({\bf p, p'})$ is symmetric in ${\bf p}$ and ${\bf p'}$. It therefore has the decomposition
$W({\bf p, p'})= \sum_{\nu} \lambda_{\nu}^{} w_{\nu}^{}({\bf p})w_{\nu}^{\ast}({\bf p'})$ where $w$'s are orthogonal,   $\Omega^{-1}\sum_{\bf p} w^{}_{\nu}({\bf p})w^{\ast}_{\nu'}({\bf p})\propto \delta_{\nu\nu'}$.  Eq.(\ref{Vqp}) also implies that  $W({\bf p, p'})$ is odd in ${\bf p}$, hence $w_{\nu}({\bf p}) = - w_{\nu}({\bf -p})$.  We shall assume $W$ has at least one positive eigenvalue, i.e. $- W$ has an attractive component. 
Finally, we shall assume that the eigenfunctions $w$ decrease sufficiently fast at large wavevector so that $I_{\nu}^{} \equiv \Omega^{-1} \sum_{\bf p}^{} |w_{\nu}({\bf p})|^2/E_{\bf p}^{}$ converges\cite{true}. 

{\bf B2. A new ground state:} Our Hamiltonian is now ${\cal H} = H_{BCS} + V_{qp}^{}$.  Note that $V_{qp}^{}$ conserves polarization $M \equiv 
N_{\uparrow}^{}- N_{\downarrow}^{}$. Since 
${\cal H}$ (for  quasi-particles) has the same form as the pairing Hamiltonian for fermions,  an attractive $W$ will augment the ``vaccum" (i.e.  $|BCS\rangle$) with quasi-particles pair fluctuations, which, as discussed in Section {\bf A},  are triplet fluctuations. 
A non-zero $h$ will then produce more  $\uparrow\uparrow$ than $\downarrow\downarrow$ pairs and hence a non-zero polarization $M$.  

There are, however,  major differences between pairing of fermions and pairing of quasiparticles. 
Since the spectra of fermions ($\xi_{\bf k}^{}$)  are gapless and those of 
quasiparticle ($E_{\bf k}^{}$) are gapped, pairing has much more dramatic effect on the former. 
Furthermore, in the absence of pairing, the ground state of fermions consists large number of particles (normal Fermi sea) whereas that of ${\cal H}$ (i.e. BCS ``vacuum"  $|BCS\rangle$) consists of  {\em no} quasiparticles.  Thus, the re-organization of normal Fermi sea caused by fermion pairing is 
much more dramatic than that of the $|BCS\rangle$ caused by quasiparticle pairing. 
Nevertheless, such reorganization is sufficient to cause the system a non-zero response to $h$, a property that does not exist in the BCS state.  

The analogy with BCS state also shows that the true ground state (with $M$ at  $h< \Delta$)  can not be obtained by performing perturbation of the spin conserving $V_{qp}$ on $|BCS\rangle$, which preserves the zero polarization of the BCS state despite $h\neq 0$.
What is needed is to go to the {\em grand canonical ensemble in spin space}, allowing  fluctuations in spin polarization, and replacing ${\cal H}$ by the mean field Hamiltonian
\begin{equation} 
{\cal H} = H_{BCS} + \frac{1}{2} \sum_{\bf k}  \left[\zeta_{\bf k}^{} (\alpha^{}_{\bf -k}
\alpha^{}_{\bf k} - \beta^{\dagger}_{\bf k}\beta_{\bf -k}^{\dagger}) + h.c\right]. 
\label{Hqp}\end{equation}
with mean field  
\begin{equation}
\zeta_{\bf p} = - \frac{2}{ \Omega}\sum_{\bf p'}W({\bf p,p'})\langle \Gamma_{\bf p'}\rangle.
\label{mean} \end{equation} 
Spin symmetry is now broken. It is then possible for the system to produce magnetization to gain energy from $h$. Eq.(\ref{Hqp}) can be diagonalized as  ${\cal H} =\sum_{\bf p}^{}  \left(   {\cal E}_{\bf p}^{(-)} A^{\dagger}_{\bf p}A^{}_{\bf p}  + 
{\cal E}_{\bf p}^{(+)} B^{\dagger}_{\bf p}B^{}_{\bf p} \right) +  E_{G}$,  
with ground state 
\begin{equation}
 |G\rangle = \prod_{\bf p} \left( \overline{u}_{\bf p} + \overline{v}_{\bf p} \alpha^{\dagger}_{\bf -p}
 \alpha^{\dagger}_{\bf p}\right)\left( \overline{\overline{u}}_{\bf p} +  \overline{\overline{v}}_{\bf p} \beta^{\dagger}_{\bf -p}
 \beta^{\dagger}_{\bf p}\right)|BCS\rangle, 
\label{HZ} \end{equation}
where $\overline{u}_{\bf k}^{}, \overline{v}_{\bf k}^{}, \overline{\overline{u}}_{\bf k}, \overline{\overline{v}}_{\bf k}$ are coherence factors\cite{comment2};   $E_{G} =  \sum_{\bf p} \left( \xi_{\bf p}^{} - 
[ {\cal E}_{\bf p}^{(-)}  + {\cal E}_{\bf p}^{(+)}]/2  \right)$ is the energy of $|G\rangle$ and 
\begin{equation}
E_{G}  - E_{BCS}^{} =  \sum_{\bf p} \left( E_{\bf p}^{} - 
[ {\cal E}_{\bf p}^{(-)}  + {\cal E}_{\bf p}^{(+)}]/2  \right) <0. 
\label{HZenergy}\end{equation}
The operators $A_{\bf p}^{} = \overline{u}_{\bf p}^{} \alpha_{\bf p}^{} +  \overline{v}_{\bf p}^{} \alpha_{\bf -p}^{\dagger}$, $ B_{\bf p}^{} = \overline{\overline{u}}_{\bf p}^{} \beta_{\bf p}^{} +  \overline{\overline{v}}_{\bf p}^{} \beta_{\bf -p}^{\dagger}$ are the new quasi-particles with energies  
\begin{equation}
 {\cal E}_{\bf p}^{(\sigma)} = \sqrt{ (E_{\bf p}^{}+ \sigma h)^2 + |\zeta_{\bf p}^{}|^2 }, \,\,\,\,\,\,\, \sigma = \pm. 
 \label{qpenergy} \end{equation}
Eq.(\ref{HZenergy}) and (\ref{qpenergy}) show that spin fluctuations $\zeta_{\bf k}^{}$ increase the energy of the excitations but decrease the energy of the ground state. They also lead to a ``magnetization" 
$m = \Omega^{-1}\sum_{\bf k} \langle c^{\dagger}_{\bf k\uparrow}  c^{}_{\bf k\uparrow} - c^{\dagger}_{\bf -k\downarrow}  c^{}_{\bf -k\downarrow} \rangle = \sum_{\bf k}^{} (\overline{v}^2_{\bf k}  - \overline{\overline{v}}^2_{\bf k})$, 
\begin{equation}
m =
\int \frac{ {\rm d}^{3}_{}k}{(2\pi)^3}
\sum^{}_{\sigma = \pm}  \left( \frac{ \sigma (E^{}_{\bf k} + \sigma h)   }{2 \sqrt{ (E^{}_{\bf k} + \sigma  h)^2  + |\zeta^{}_{\bf k}|^2} }\right).
\label{m} \end{equation}

{\bf B3. Triplet mean field and magnetization $m$:} 
Evaluating $\langle \Gamma_{\bf p'}^{}\rangle$ in eq.(\ref{mean}), we have 
\begin{equation}
\zeta_{\bf p} = \frac{1}{\Omega} \sum_{\bf p', \sigma = \pm } W({\bf p, p'}) \left( 
\frac{\zeta_{\bf p'}}{ \sqrt{(E_{\bf p'}+\sigma h)^2  + |\zeta_{\bf p'}|^2 }} \right).
\label{zeta} \end{equation}
If  $\lambda_{\nu_{o}}$ is the largest positive eigenvalue of $W$, 
eq.(\ref{zeta}) then has a solution $\zeta_{\bf p} = w_{\nu_{o}}({\bf p}) {\cal Q}_{\nu_{o}}$, with 
\begin{equation}
 {\cal Q}_{\nu_{o}} =\lambda_{\nu_{o}}  \sum_{ \sigma=\pm} \int \frac{ {\rm d}^{3}k}{(2\pi)^3}
  \frac{ |w_{\nu_{o}}({\bf k})|^2  
  {\cal Q}_{\nu_{o}}}
{\sqrt{(E_{\bf k} + \sigma h)^2  +  |w_{\nu_{o}}({\bf k})|^2   |{\cal Q}_{\nu_{o}}|^2 }}. 
\label{Q}  \end{equation}
It is useful to compare eq.(\ref{Q}) with eq.(\ref{BCSgap}).  Since $\xi_{\bf k}^{}$ is gapless, 
the integral in eq.(\ref{BCSgap})  is  logarithmically divergent as $\Delta \rightarrow 0$. As a result, eq.(\ref{BCSgap}) 
can always be satisfied (for all $g$) by an appropriate choice of $\Delta$.  
Eq.(\ref{Q}) is different.  Due to the gap in $E_{\bf p}$,  the integral
in eq.(\ref{Q}) at $h=0$ is bounded by the convergent integral $I_{\nu_{o}}^{} = \Omega^{-1}\sum_{\bf p}^{} | w_{\nu_{o}}({\bf p})|^2/E_{\bf p}^{}$. 
We then have the following possibilities : 

\noindent {\bf (i)}  For  $\lambda_{\nu_{o}}> \lambda^{}_{c} \equiv 1/I_{\nu_{o}}^{}$, eq.(\ref{Q}) implies
${\cal Q}_{\nu_{o}}^{} = C + D h^2+ .. $ as $h\rightarrow 0$,  where $C$ and $D$ are constants. 
In other words, for sufficiently strong interactions, triplet fluctuations ($C$) can exist in the 
 the ground state of equal spin population.  Eq.(\ref{m}) then implies 
$m = M/\Omega=\chi h + O(h^{2})$, where $\chi$ is the susceptibility
\begin{equation}
\chi(\mu, \Delta) = \frac{1}{\Omega}\sum_{\bf k}  \frac{ |\zeta^{o}_{\bf k}|^2  }{ (E^{2}_{\bf k} + |\zeta^{o}_{\bf k}|^2 )^{3/2}   } , 
\end{equation}
and $\zeta^{o}_{\bf k} = w_{\nu_{o}}({\bf k}) {\cal Q}_{\nu_{o}}^{o}$ is the solution of eq.(\ref{Q}) at $h=0$.  

\noindent {\bf (ii)} For  $\lambda_{\nu_{o}}< \lambda_{c}^{}$,  ${\cal Q}_{\nu_{o}}$ vanishes at $h=0$.  However, since the integral in eq.(\ref{Q}) is logarithmicaly divergent as $h\rightarrow \Delta$ and ${\cal Q}_{\nu_{o}}\rightarrow 0$,  a nonzero  ${\cal Q}_{\nu_{o}}$ can always be found by increasing $h$ {\em before it reaches} $\Delta$. The critical $h$ (denoted as $h_{c}^{}$) when ${\cal Q}_{\nu_{o}}$ becomes non-zero is given by  $\lambda_{\nu_{o}}^{-1} = \Omega^{-1}\sum_{\bf k, \sigma = \pm }^{} |w_{\nu_{o}}({\bf k})|^2
/(E_{\bf k}^{} + \sigma h_{c}^{})$.  As $h\rightarrow h_{c}^{}$ from above, we have ${\cal Q}_{\nu_{o}}^{}\propto (h-h^{}_{c})^{1/2}$, $m(h) \propto (h-h^{}_{c})^{}$.  

\noindent {\bf (iii)} At $\lambda = \lambda_{c}$, we have ${\cal Q}_{\nu_{o}}^{}\sim h$ and $m\sim h^{3}$.



At unitarity, where the only energy scale at $T=0$ is  $\mu$ (or Fermi energy $E_{F}$), the system acquires a universal thermodynamics\cite{Houni} with magnetization given by $m(\mu) = n(\mu) G(h/\mu)$, where $n(\mu) \propto \mu^{3/2}$ is the number density, and $G(h/\mu)$ is a dimensionless universal function.  There is no need to consider the $\Delta$-dependence of $m$, since at unitarity we have $\Delta = \alpha \mu$, where $\alpha$ is a universal constant, ($\alpha= 1.16$ in BCS theory, and 1.22 according to Quantum Monte Carlo calculations\cite{Monte}).  For large quasiparticle interactions, $\lambda>\lambda_{c}^{}$, $m$ is linear $h$. This implies  $G(h/\mu) \propto h/\mu$, or $m(\mu) \propto \mu^{1/2} h$. In the presence of a trap $V({\bf r})$ (which vanishes at, say, ${\bf r}=0$), local density approximation ($\mu \rightarrow \mu - V({\bf r})$) implies 
\begin{equation}
m({\bf r}) = m({\bf 0}) ( 1 - V({\bf r})/\mu)^{1/2} 
\label{profile} \end{equation}
in the region where $\mu-V({\bf r})>0$, a property that can be tested experimentally. For weak quasi-particle interactions so that $G(h/\mu)$ is not linear in $h$ for small $h$, the full scaling form  $m(\mu) = n(\mu) G(h/\mu)$ has to be used to determine the spin density profile. 

\begin{figure}[bp]
\begin{center}
\includegraphics[width=7.0cm]
{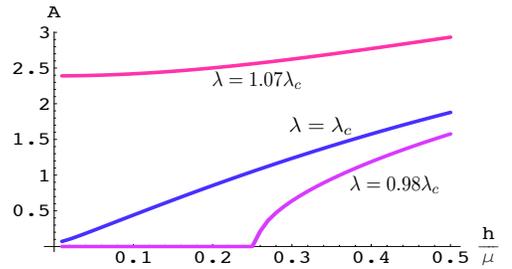}\caption{ The triplet fluctuation $A$ (in eq.(\ref{A})) as  a function of $h/\mu$. For $\lambda > \lambda_{c}^{}$, $\lambda = \lambda_{c}^{}$, and $\lambda < \lambda_{c}^{}$, we have $A\propto$ constant, 
$A\propto h$ and $A\propto \sqrt{h-h_{c}}$ respectively. \label{A_h}}
\end{center}
\end{figure}

{\bf C. A specific model of quasi-particle interaction at unitarity:}  In principle, the quasi-particle interaction is determined by the vertex function $\Gamma(p_{1}, p_{2}; p_{3}, p_{4})$.   As in studying fermion pairing, one needs to first identify the relevant interactions. Since spin fluctuations are equivalent to 
quasi-particle pairing (see Sec. {\bf A}),  we look for terms in the interactions that generate triplet pairs with zero total momentum in the particle-hole channel\cite{particlehole}, which are of the form
$\overline{V}_{qp} =  \Omega^{-1} \sum_{\bf p, p'} F({\bf p, p'}) c^{\dagger}_{\bf p'\uparrow} c^{\dagger}_{\bf p\downarrow} c^{}_{\bf p'\downarrow}c^{}_{\bf p\uparrow} $,
where $F({\bf p, p'})$ an interaction function symmetric in ${\bf p}$ and ${\bf p'}$. 
 Expressing the $c$'s 
 in terms of $\alpha_{\bf k}^{}$  and $\beta_{\bf k}$, we find $ \overline{V_{qp}^{}}  = V_{qp}^{}+ U_{qp}$ where  $V_{qp}^{}$ is precisely eq.(\ref{Vqp}) with  $W({\bf p, p'}) = F({\bf p,p'})\chi_{\bf p}^{} \chi^{}_{\bf p'}$, with $\chi_{\bf p} = u_{\bf p}^{}v_{\bf p}^{}$.  The $ U_{qp}$ are terms that do not produce triplet fluctuations and will be ignored\cite{morecomment}.

\begin{figure}[tbp]
\begin{center}
\includegraphics[width=7.0cm]
{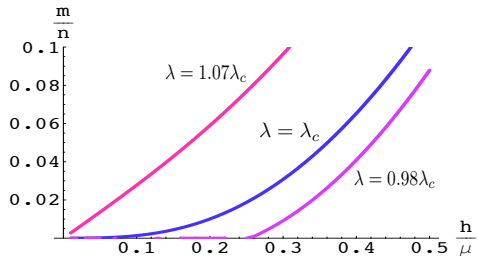}\caption{ The polarization $m/n$ as a function of $h/\mu$ according to eq.(\ref{m}). For $\lambda > \lambda_{c}^{}$, $\lambda = \lambda_{c}^{}$, and $\lambda < \lambda_{c}^{}$, we have $m\propto h$,  
$m \propto h^3$ and $m \propto (h-h_{c})$ respectively.\label{M}  }
\end{center}
\end{figure}
 
Still, there is the problem of finding the expression of $F({\bf p, p'})$. Once again, simplifications can be obtain at unitarity, since the only momentum scale at $T=0$ is $k_{o}^{}= \sqrt{2M\mu/\hbar^2}$.  If we further assume that the existence of an effective field theory for the fermions where the interaction coefficients can be expanded in powers of ${\bf p}$, then in the lowest order in momentum, we have 
\begin{equation}
W({\bf p,p'})=  F({\bf p,p'})\chi_{\bf p}^{} \chi^{}_{\bf p'} = \frac{4\pi\hbar^2 \lambda {\bf p}\cdot {\bf p'}}{M k_{o}^{3}} \chi^{}_{\bf p}\chi^{}_{\bf p'}.
\end{equation}
where $\lambda$ is dimensionless. 
Eq.(\ref{zeta}) then has a solution 
$\zeta_{\bf k}^{} = \mu \chi_{\bf k}^{}{\bf p}\cdot {\bf A}/k_{o}^{}$, with 
\begin{equation}
 {\bf A} = \frac{4 \pi \lambda \hbar^2}{M k^{3}_{o}  }\frac{1}{\Omega} \sum_{ {\bf k} , \sigma = \pm}^{} 
 \frac{  (\chi^{2}_{\bf k} {\bf k}) ( {\bf k}\cdot {\bf A}^{})}{2\sqrt{ (E_{\bf k}^{}-\sigma h)^2 + 
   |\chi_{\bf k}^{}{\bf k}\cdot {\bf A}|^2 }}  
\label{A} \end{equation}
As discussed in Sec.{\bf B2}, the solution of eq.(\ref{A}) depends on whether $\lambda> \lambda_{c}^{}$, where $\lambda_{c}^{-1} = \frac{4 \pi \lambda \hbar^2}{M k^{3}_{o} \Omega } \sum_{\bf k}^{} 
 \frac{  (\chi^{2}_{\bf k} {\bf k}) ( {\bf k}\cdot {\bf A}^{})}{ E_{\bf k}^{} }$.   Taking the BCS result $\Delta = 1.16\mu$ at unitarity,  we have  $\lambda_{c}^{}= 2.14$.  We also find that the solution of eq.(\ref{A}) with lowest energy is ${\bf A} = A (\hat{\bf x} \pm i \hat{\bf y})/\sqrt{2}$.  
 
\begin{figure}[htbp]
\begin{center}
\includegraphics[width=8.0cm]
{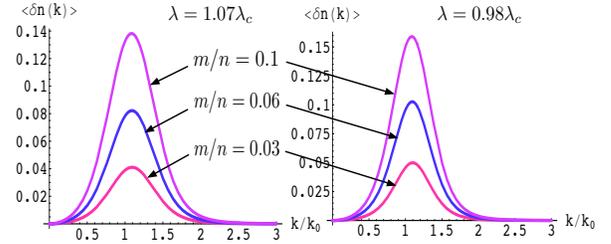}\caption{ The difference of angular averaged momentum distribution between different spin populations, $\delta n({\bf k})$.  \label{momentum} }
\end{center}
\end{figure}

 The behavior of triplet fluctuation $A$ and polarization $m/n$ as a function of $h/\mu$ for different 
 $\lambda$ are shown in Fig.\ref{A_h} and \ref{M}, which illustrate the behaviors discussed in {\bf (i)} to {\bf (iii)} mentioned in Sec.{\bf B3}.  (Note that $A \propto {\cal Q}_{\nu_{o}}^{}$.)  
Fig.\ref{momentum} shows the difference between momentum distributions of different spin components, averaged
over all angles,  $\delta n(k) = (4\pi)^{-1}\int {\rm d}\hat{\bf k} [n_{\uparrow}^{}({\bf k}) - n_{\downarrow}^{}({\bf k})]$ for different quasiparticle interaction $\lambda$ and different polarization 
$m/n$.  It is interesting to note that the difference expands over  a wide range of momenta around $k_{o}$.  Moreover, the difference is not very sensitive to whether $\lambda>\lambda_c$ or   $\lambda<\lambda_c$. For $m/n = 0.1$, the difference at $k_{o}$ can be as large are 0.14.  Measurement of momentum distribution can therefore be used to detect the existence of this new state eq.(\ref{HZ}). 
In addition to the profile for magnetization (eq.(\ref{profile})) and $\delta n(k)$, the new state eq.(\ref{HZ}) can also be detected by noise measurements.  Since eq.(\ref{HZ}) contains $\uparrow\uparrow$ and $\downarrow\downarrow$ pairs, detection of correlations between like spin fermions with {\em opposite momenta} at small polarization will make a strong case for the new state eq.(\ref{HZ}).

{\bf  Concluding remarks:}  
We have shown that quasiparticle interactions can introduce triplet fluctuations in the BCS state. The resulting state eq.(\ref{HZ})  will have non-zero ``spin susceptibility" even at very low ``magnetic fields" $h$. Although our calculations are based on a specific model on the quasiparticle interaction, many consequences of these interactions can be deduced on general grounds.  To compare energies with phase separation state requires a better understanding of the vertex function $\Gamma(p_{1}, p_{2}; p_{3}, p_{4})$ at unitarity, which is an intriguing and  challenging problem in itself and will be studied elsewhere.  In any case, the new state eq.(\ref{HZ}) has many unique features in the density profile (eq.(\ref{profile})), in the momentum distribution (figure 3),  and noise correlation between like spins. All these properties can be measured experimentally. 

TLH would like to thank Randy Hulet for many stimulating discussions. This work is supported by  NASA GRANT-NAG8-1765  and NSF Grant DMR-0426149.


\begin{thebibliography}{99}
\bibitem{morecomment} $U_{qp}^{}$ can be considered as the residual interactions for ${\cal H}$. These interactions, however, do not change the triplet fluctuations in $|G\rangle$.   Note also that only terms in $F({\bf p, p'})$ that are odd in ${\bf p}$ and ${\bf p'}$ will present in $V_{qp}^{}$.
\end{thebibliography}
\end{document}